\RequirePackage[svgnames]{xcolor}

\documentclass[11pt,letterpaper]{mystyle}

\usepackage[all]{hypcap}
\usepackage[svgnames]{xcolor}
\usepackage[comma,authoryear,compress]{natbib}
\usepackage{hyperref}[citecolor=lightblue]

\hypersetup{
    colorlinks = true,
    citecolor = {YaleBlue},
}

\usepackage{tcolorbox} 

\usepackage{arydshln}
\usepackage{multirow}
\usepackage{algorithm}
\usepackage{algorithmicx}
\usepackage{algpseudocode}
\usepackage{microtype}
\usepackage{graphicx}
\expandafter\def\csname ver@subfig.sty\endcsname{}
\usepackage{booktabs} %
\usepackage{float}
\usepackage{bigstrut}

\usepackage{amsmath}
\usepackage{amssymb}
\usepackage{mathtools}
\usepackage{amsthm}
\usepackage{mathrsfs}
\usepackage{nicefrac}
\usepackage{dsfont}
\usepackage{enumitem}
\usepackage{subcaption}
\usepackage{graphicx,subfig}
\usepackage{cleveref}
\usepackage{bxcoloremoji}

\usepackage{float}

\usepackage[utf8]{inputenc} %
\usepackage[T1]{fontenc}    %
\usepackage{hyperref}       %
\usepackage{url}            %
\usepackage{booktabs}       %
\usepackage{amsfonts}       %
\usepackage{nicefrac}       %
\usepackage{microtype}      %
\usepackage{graphicx}
\usepackage{subcaption} 
\usepackage{amssymb}
\usepackage{fdsymbol}
\usepackage{wrapfig}
\usepackage{lipsum}
\usepackage{enumitem}
\usepackage{stackengine}
\usepackage[font=small,labelfont=bf]{caption}
\usepackage{color}
\usepackage{adjustbox}

\usepackage{rotating}
\usepackage{makecell}

\newcommand{\x}{\mathbf{x}}

\definecolor{blanchedalmond}{rgb}{1.0, 0.92, 0.8}
\definecolor{carmine}{rgb}{0.59, 0.0, 0.09}
\definecolor{lightblue}{rgb}{0.22,0.45,0.70}%

\renewcommand{\mathbf}{\boldsymbol}

\makeatletter
\def\Ddots{\mathinner{\mkern1mu\raise\p@
\vbox{\kern7\p@\hbox{.}}\mkern2mu
\raise4\p@\hbox{.}\mkern2mu\raise7\p@\hbox{.}\mkern1mu}}
\makeatother

\definecolor{amaranth}{rgb}{0.9, 0.17, 0.31}
\definecolor{antiquebrass}{rgb}{0.8, 0.58, 0.46}
\definecolor{antiquefuchsia}{rgb}{0.57, 0.36, 0.51}
\definecolor{chromeyellow}{rgb}{0.31, 0.47, 0.26}

\newcommand{\paperlogo}{\raisebox{-1.5pt}{\includegraphics[height=2.05em]{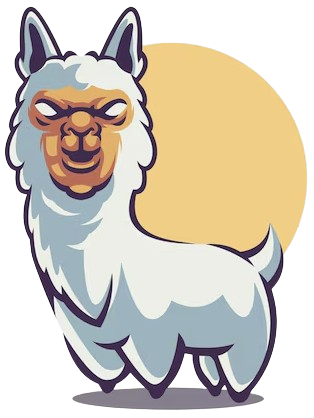}}}

\newtcolorbox{AIbox}[2][]{aibox,title=#2,#1}
\definecolor{lightblue}{rgb}{0.22,0.45,0.70}%
\definecolor{Gray}{gray}{0.95}
\definecolor{Cornsilk}{rgb}{1.0, 0.97, 0.86}

\usepackage{amsmath}

\usepackage[all]{hypcap}

\title{\paperlogo{} Audio-Thinker:  Guiding Audio Language Model When and How to Think via Reinforcement Learning}

\runningtitle{\paperlogo{} Audio-Thinker:  Guiding Audio Language Model When to Think and How to Think via Reinforcement Learning}

\author{%
  Shu Wu$^{1}$ \quad Chenxing Li$^{1}$ \quad Wenfu Wang$^{1}$ 
  \\ 
  \textbf{Hao Zhang}$^{2}$ \quad \textbf{ Hualei Wang}$^{1}$ \quad \textbf{Meng Yu}$^{2}$ \quad \textbf{Dong Yu}$^{2}$
}
\affil[1]{Tencent AI Lab, Beijing }
\affil[2]{Tencent AI Lab, Seattle}

\begin{document}

\begin{abstract}
Recent advancements in large language models, multimodal large language models, and large audio language models (LALMs) have significantly improved their reasoning capabilities through reinforcement learning with rule-based rewards. However, the explicit reasoning process has yet to show significant benefits for audio question answering, and effectively leveraging deep reasoning remains an open challenge, with LALMs still falling short of human-level auditory-language reasoning.
To address these limitations, we propose Audio-Thinker, a reinforcement learning framework designed to enhance the reasoning capabilities of LALMs, with a focus on improving adaptability, consistency, and effectiveness. Our approach introduces an adaptive think accuracy reward, enabling the model to adjust its reasoning strategies based on task complexity dynamically. Furthermore, we incorporate an external reward model to evaluate the overall consistency and quality of the reasoning process, complemented by think-based rewards that help the model distinguish between valid and flawed reasoning paths during training. Experimental results demonstrate that our Audio-Thinker model outperforms existing reasoning-oriented LALMs across various benchmark tasks, exhibiting superior reasoning and generalization capabilities.

\vspace{2mm}

\textit{Keywords: Large Audio Language Model, Multimodal Reasoning, Reinforcement Learning, Adaptive Thinking}

\vspace{5mm}

\coloremojicode{1F4C5} \textbf{Date}: August 9, 2025

\coloremojicode{1F4E7} \textbf{Contact}: \href{mailto:shoookwu@outlook.com}{shoookwu@outlook.com} \quad \href{mailto:chenxingli@tencent.com}{chenxingli@tencent.com} \quad \href{mailto:dyu@global.tencent.com}{dyu@global.tencent.com}

\end{abstract}

\maketitle
\vspace{3mm}
\vspace{-4mm}
\begin{figure*}[h]
    \centering
    \includegraphics[width=0.9\textwidth]{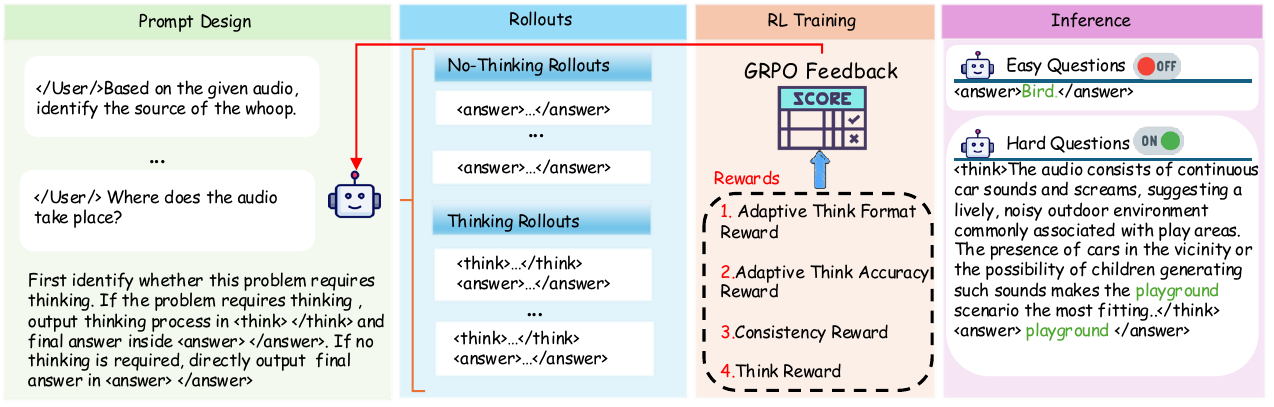} 
    
    \caption{\textbf{Overview of the Audio-Thinker framework.} As illustrated in the block Inference, the LALMs trained using the Audio-Thinker framework are capable of achieving adaptive reasoning capabilities that scale according to the complexity and difficulty of the given task.}
    \label{overview}
    \end{figure*}   
\section{Introduction}

Recent advancements in large language models (LLMs) demonstrate that reasoning can be significantly enhanced through techniques such as chain-of-thought prompting, diverse cognitive frameworks, and reinforcement learning (RL). RL-tuned models excel in complex tasks, including math problem-solving and coding, with strategies like GRPO providing substantial improvements beyond traditional supervised learning methods. Research reveals that smaller models tend to thrive with structured thinking, while larger models perform better with unstructured approaches.

Recent studies \cite{huang2025visionr1incentivizingreasoningcapability,liu2025visualrftvisualreinforcementfinetuning,pan2025medvlmr1incentivizingmedicalreasoning,zhou2025r1zerosahamomentvisual} have advanced RL techniques in Multimodal Large Language Models (MLLMs) across domains like object recognition \cite{liu2025visualrftvisualreinforcementfinetuning}, semantic segmentation \cite{liu2025segzeroreasoningchainguidedsegmentation}, and video analysis \cite{sun2025videosalmonno1reasoningenhancedaudiovisuallarge}. These methods enhance MLLM capabilities, especially in data-scarce scenarios, achieving SFT-level performance in in-domain tasks and outperforming SFT in out-of-distribution (OOD) evaluations.

The realm of audio-language reasoning and reinforcement learning fine-tuning (RLF) remains relatively uncharted. Prominent Large Audio-Language Models (LALMs) such as Audio Flamingo \cite{kong2024audio}, SALMONN \cite{tang2023salmonn}, and Qwen2-Audio \cite{qwen2} have significantly advanced audio comprehension in various benchmarks. However, these models primarily concentrate on perception and basic question-answering tasks without incorporating explicit reasoning processes. Subsequently, Audio-Reasoner \cite{xie2025audio} employed a structured reasoning methodology on Qwen2-Audio, while R1-AQA \cite{li2025reinforcement} implemented the GRPO algorithm, discovering that merely adding a reasoning chain does not yield substantial improvements. In contrast, SARI \cite{wen2025saristructuredaudioreasoning} fine-tunes Qwen2.5-Omni using reinforcement learning in tandem with both structured and unstructured reasoning. However, its performance does not match that of Omni-R1 \cite{rouditchenko2025omnir1reallyneedaudio}, which is trained exclusively with reinforcement learning. This highlights the ongoing challenge of effectively leveraging reinforcement learning to enhance reasoning capabilities in audio question-answering tasks.

In this study, we address the challenge by introducing a reinforcement learning framework known as Audio-Thinker, designed to enhance the adaptive, consistent, and effective reasoning capabilities of LALMs. Audio-Thinker employs an adaptive thinking mode policy that determines when the model should engage in ``thinking'', based on the complexity of the query. Moreover, it integrates an external expert LLM to provide thought-based supervision, guiding the model in generating coherent and effective reasoning processes.
The main contributions are as follows. 

\begin{itemize}
\item \textbf{Audio-Thinker:} We present Audio-Thinker, a universal reinforcement learning framework that empowers LALMs to explore effective reasoning policies while simultaneously enhancing reasoning quality.
\item \textbf{When to Think:} We introduce an adaptive thinking accuracy reward that trains LALMs to modulate their reasoning strategies according to task complexity, directing the model to find optimal reasoning approaches.
\item \textbf{How to Think:} We integrate think-based rewards that evaluate the consistency and quality of reasoning, allowing the model to distinguish between sound and flawed reasoning processes during training.
\item \textbf{State-of-the-Art Performance:}
In the experiments, our Audio-Thinker models consistently outperform existing LALMs on diverse benchmarks, including MMAU \cite{mmau2024}, MMAR \cite{ma2025mmarchallengingbenchmarkdeep}, and AIR \cite{yang2024airbenchbenchmarkinglargeaudiolanguage}, highlighting its strong reasoning and generalization abilities. 
\end{itemize}

\label{sec:intro}
\vspace{-1mm}

\section{Relate Works}
\label{sec:related_work}

\subsection{Large Audio Language Models}
The rapid advancement of LLMs catalyzes the evolution of MLLMs, which possess the capacity to comprehend and reason across a diverse array of data modalities, including auditory information. Exemplary instances of LALMs, such as Qwen2-Audio \cite{qwen2}, Audio Flamingo \cite{kong2024audio}, and SALMONN \cite{tang2023salmonn}, exhibit remarkable capabilities in audio understanding and processing.

\subsection{Language and Multimodal Reasoning}
Recently, models such as OpenAI-o1 \cite{jaech2024openai}, Kimi K1.5 \cite{team2025kimi}, and DeepSeekR1 \cite{guo2025deepseek} draw attention for enhancing reasoning performance through reinforcement learning \cite{jin2025search,peng2025lmm,openr1}. This progress spurs follow-up research, including successful method replications \cite{xie2025logic} and efforts to improve algorithmic efficiency \cite{yu2025dapo}. Reinforcement learning is increasingly applied to vision-language models \cite{yang2025r1,feng2025video,huang2025vision}. For instance, Vision-R1 \cite{huang2025vision} proposes Progressive Thinking Suppression Training to reduce overthinking, Video-R1 \cite{feng2025video} explores R1-style reinforcement learning for video reasoning, and LMM-R1 introduces a rule-based RL framework to advance multimodal reasoning.

\subsection{Audio Models with Reasoning}
Recent efforts concentrate on enhancing reasoning capabilities in audio-language models. A notable example is Mellow \cite{mellow}, a lightweight audio-language model that demonstrates exceptional reasoning abilities. Despite having only 167 million parameters and being trained on 1.24 million examples, Mellow outperforms larger State-of-the-Art Performance (SOTA) models across various domains. Audio-CoT \cite{ma2025audio} is the first model to explore Chain-of-Thought (CoT) reasoning in audio-language models; however, it does not incorporate model updates and offers limited advancements for tackling complex issues. Additionally, another significant model, Audio-Reasoner \cite{xie2025audio}, is specifically designed for deep reasoning in audio tasks. This model introduces a structured reasoning process that utilizes a large-scale dataset (CoTA) and employs a multi-phase "thinking" architecture comprising planning, captioning, reasoning, and summarization before generating its final response. Furthermore, R1-AQA \cite{li2025reinforcement} utilizes the GRPO algorithm to fine-tune the Qwen2-Audio model for audio question-answering tasks, enhancing reasoning accuracy with less data through reward-driven optimization. Concurrently, SARI \cite{wen2025saristructuredaudioreasoning} fine-tunes Qwen2.5-Omni \cite{xu2025qwen25omnitechnicalreport} using reinforcement learning, presenting a study focused on improving the reasoning capabilities of audio multimodal models by leveraging explicit CoT training and curriculum-guided reinforcement learning. Finally, Omni-R1\cite{rouditchenko2025omnir1reallyneedaudio} fine-tunes Qwen2.5-Omni with GRPO, employing a straightforward yet effective prompt that streamlines training and testing, ultimately achieving a new SOTA performance.

\section{Observations and Motivations}
\subsection{O1: Explicit Thinking Does Not Always Yield Effective Results.}
Research on LLMs and MLLMs frequently posits that explicit reasoning can bolster reasoning capabilities. However, investigations conducted by R1-AQA and Omni-R1 reveal that the explicit reasoning process has not yielded substantial advantages for Automated Question Answering (AQA) tasks. Thus, how to effectively leverage \textit{\textbf{deep thinking}} remains an open challenge for future work.

\subsection{O2: Prompting Alone Does Not Enable Adaptive Thinking}

One possible solution to the issue identified in O1 is the implementation of \textbf{\textit{adaptive thinking}} \cite{zhang2025adaptthinkreasoningmodelslearn,li2025thinkthinkstudyexplicit}, whereby the model dynamically determines whether reasoning is warranted based on input characteristics. This can be achieved through a prompting strategy that enables context-aware adaptation to question complexity.

\begin{figure}[t]
    \centering
    \includegraphics[width=0.6\columnwidth]{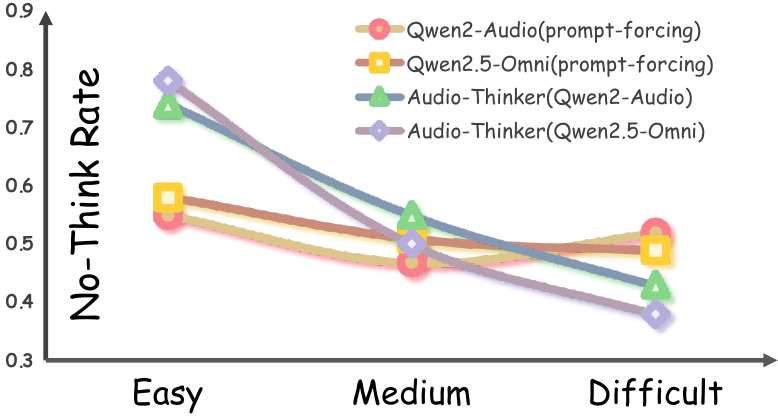} 
    
    \caption{No-Thinking Rate by Difficulty on MMAU-test-mini. Prompt-forcing models show a flat distribution, indicating no sensitivity to problem complexity, while Audio-Thinker models exhibits a clear trend, demonstrating difficulty-aware reasoning.}
    \label{rate}
    \end{figure}  

To evaluate performance, we use a prompt strategy (see Figure \ref{overview}, Block ``Prompt Design'') and assess results on the MMAU-test-mini dataset. As shown in Figure \ref{rate}, we analyze the ``no-thinking'' rate across three complexity levels. Notably, prompt-forced models show no clear trend, indicating their reasoning activation is largely insensitive to problem difficulty. This suggests limited adaptability in deciding when deep thinking is needed.

\subsection{Guiding LALMs When and How to Think}
Based on current observations, existing LALMs lack adaptive thinking and sufficient supervision over their reasoning processes during training, which may hinder generalization. To address this, we propose Audio-Thinker, an audio-language reinforcement learning framework that promotes difficulty-aware, consistent, and effective reasoning. As shown in Figure \ref{rate}, the model trained with Audio-Thinker demonstrates clear difficulty-aware reasoning.

\section{Audio Thinker}
\label{sec:method}

\begin{figure*}[h]
    \centering
    \includegraphics[width=0.9\textwidth]{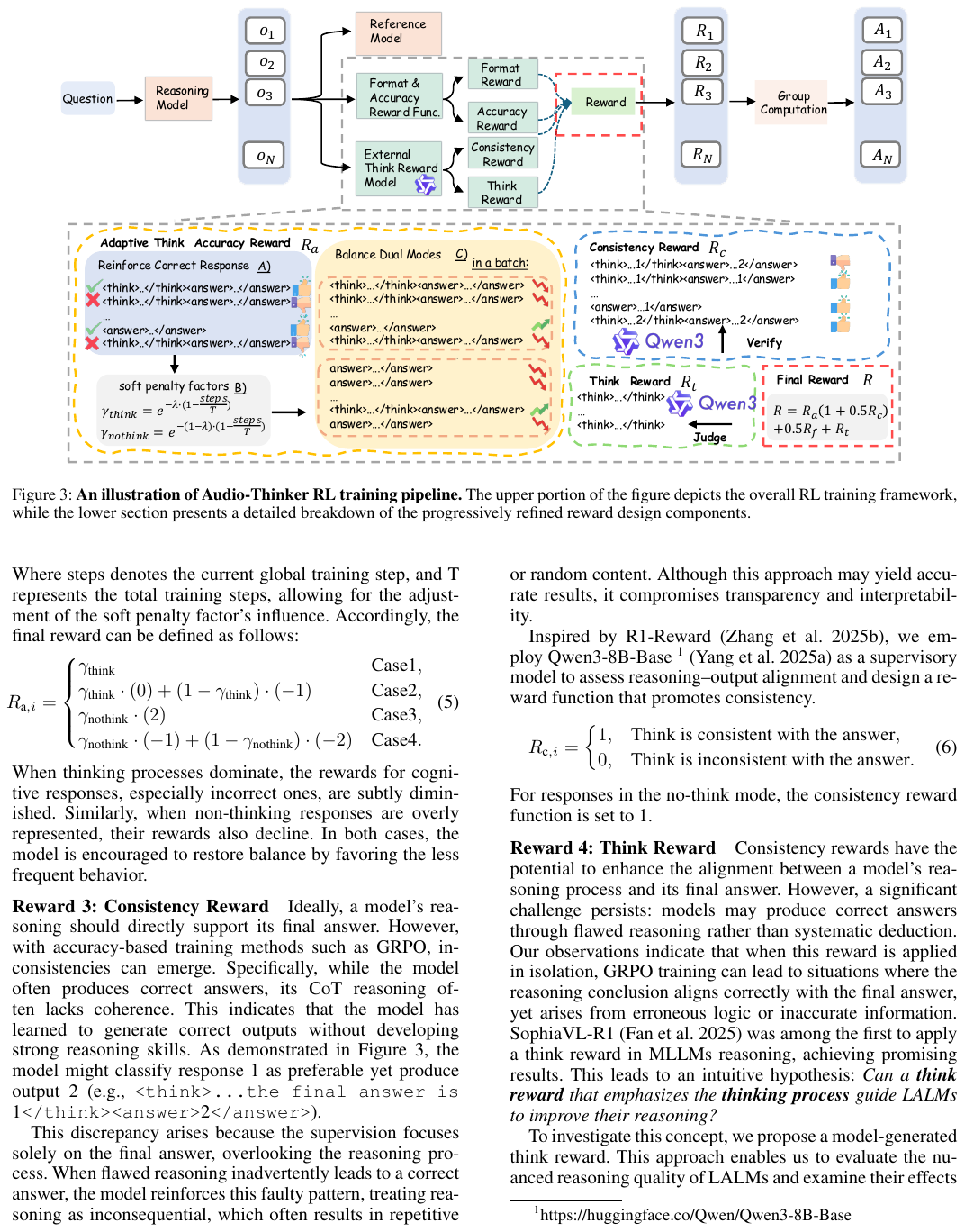}
    \caption{\textbf{An illustration of Audio-Thinker RL training pipeline.} The upper portion of the figure depicts the overall RL training framework, while the lower section presents a detailed breakdown of the progressively refined reward design components.}
    \label{detail}
\end{figure*}

 As depicted in Figure \ref{overview}, Audio-Thinker consists of two primary components:

\begin{itemize}
\item Adaptive Thinking Prompt Design: A prompting strategy that facilitates stochastic transitions between thinking and non-thinking modes in LALMs.
\item Reinforcement Learning Training Framework: As shown in Figure \ref{detail}, our approach employs a progressively refined reward function, enabling LALMs to discern the necessity of reasoning and to follow the most effective reasoning trajectory toward the solution.
\end{itemize}
Below, we provide a detailed explanation of the implementation of each module.

\subsection{Prompt Design}
We prompt the model to first assess whether a query requires reasoning, and then either generate a reasoning process if needed or provide a direct answer otherwise. Details of the prompt are provided in Appendix \ref{appendix:adaptive}.

\subsection{Progressively Refined Reward Designs}
\subsubsection{Reward 1: Adaptive Think  Format  Reward }
We prompt LALMs to decide whether reasoning is needed and then generate either a reasoned response or a direct answer accordingly (see Appendix \ref{appendix:adaptive} for the detailed prompt design). Both formats receive a format reward of 1.

\subsubsection{Reward 2: Adaptive Think  Accuracy  Reward }

As shown in Figure \ref{rate}, the prompt-only control approach has a key limitation: without feedback, the model cannot determine when reflective thinking is necessary. Inspired by AutoThink \cite{tu2025learningthinkshapingadaptive}, we propose the Adaptive Think Accuracy Reward (ATAR) to guide the model in deciding whether to engage in deep reasoning based on problem complexity, as illustrated in Figure \ref{detail}, Block ``Adaptive Think Accuracy Reward''. We assign higher rewards for correct answers that do not require reflection and impose stricter penalties for incorrect responses. We define four cases: \textbf{Case 1}: think and correct, \textbf{Case 2}: think and incorrect, \textbf{Case 3}: no-think and correct, \textbf{Case 4}: no-think and incorrect. Each sample \(i\) receives an initial reward \(R_{\text{a}, i} \in \{+1, 0, +2, -1\}\) for Cases 1, 2, 3, and 4, respectively.

This reward structure encourages difficulty-aware behavior; however, it may cause instability in the early stages of training. The model might converge on a degenerate policy, consistently choosing either to think or to skip, depending on which option seems to yield a higher expected reward in the short term. This tendency limits exploration and hampers further optimization. To mitigate this issue, we integrate the implementation of batch-level reward balancing.

Let $\lambda \in [0, 1]$ represent the proportion of Think trajectories in a training batch, with $1 - \lambda$ indicating the proportion of No-think samples. For both think and No-think samples, we calculate soft penalty factors:

\begin{equation}
\gamma_{\text{think}} = e^{-\lambda} ,
\end{equation}

\begin{equation}
\gamma_{\text{nothink}} = e^{-(1-\lambda)} .
\end{equation}

The introduction of soft penalty factors aids the model in achieving behavioral stability between thinking and non-thinking modes during the initial phases of training. However, this also constrains the model's ability to evolve freely within each mode. To address this limitation, we propose a strategy that gradually reduces the impact of soft penalty factors as training progresses. This approach encourages the reasoning model to increasingly rely on the more original and accurate rule-based rewards in the later stages, with the soft penalty factor converging towards a value of 1. The final soft penalty factors are defined as follows:
\begin{equation}
\gamma_{\text{think}} = e^{-\lambda\cdot (1-\frac{steps}{T}) } ,
\end{equation}

\begin{equation}
\gamma_{\text{nothink}} = e^{-(1-\lambda) \cdot (1-\frac{steps}{T})} .
\end{equation}
Where steps denotes the current global training step, and T represents the total training steps, allowing for the adjustment of the soft penalty factor's influence. Accordingly, the final reward can be defined as follows:
\begin{equation}
\hspace*{-10pt} %
R_{\text{a}, i} = 
\begin{cases} 
 \gamma_{\text{think}} &  \text{Case 1}, \\
 \gamma_{\text{think}} \cdot (0)+(1-\gamma_{\text{think}}) \cdot (-1) &  \text{Case 2}, \\
 \gamma_{\text{nothink}} \cdot (2) &  \text{Case 3}, \\
 \gamma_{\text{nothink}} \cdot (-1) + (1-\gamma_{\text{nothink}}) \cdot (-2) &  \text{Case 4}.
\end{cases} 
\end{equation}
When thinking processes dominate, the rewards for cognitive responses, especially incorrect ones, are subtly diminished. Similarly, when non-thinking responses are overly represented, their rewards also decline. In both cases, the model is encouraged to restore balance by favoring the less frequent behavior.

\subsubsection{Reward 3: Consistency Reward}

Ideally, a model's reasoning should directly support its final answer. However, with accuracy-based training methods such as GRPO, inconsistencies can emerge. Specifically, while the model often produces correct answers, its CoT reasoning often lacks coherence. This indicates that the model has learned to generate correct outputs without developing strong reasoning skills. As demonstrated in Figure 3, the model might classify response 1 as preferable yet produce output 2 (e.g., \texttt{<think>...the final answer is} 1\texttt{</think>}\texttt{<answer>}2\texttt{</answer>}).

This discrepancy arises because the supervision focuses solely on the final answer, overlooking the reasoning process. When flawed reasoning inadvertently leads to a correct answer, the model reinforces this faulty pattern, treating reasoning as inconsequential, which often results in repetitive or random content. Although this approach may yield accurate results, it compromises transparency and interpretability.

Inspired by R1-Reward \cite{zhang2025r1rewardtrainingmultimodalreward}, we employ Qwen3-8B-Base \footnote{\url{https://huggingface.co/Qwen/Qwen3-8B-Base} } \cite{yang2025qwen3} as a supervisory model to assess reasoning–output alignment and design a reward function that promotes consistency.
\begin{equation}
R_{\text{c}, i} = 
\begin{cases} 
1, & \text{Think is consistent with the answer}, \\
0,& \text{Think is inconsistent with the answer}. 
\end{cases}
\end{equation}
For responses in the no-think mode, the consistency reward function is set to 1.

\subsubsection{Reward 4: Think  Reward}
Consistency rewards have the potential to enhance the alignment between a model's reasoning process and its final answer. However, a significant challenge persists: models may produce correct answers through flawed reasoning rather than systematic deduction. Our observations indicate that when this reward is applied in isolation, GRPO training can lead to situations where the reasoning conclusion aligns correctly with the final answer, yet arises from erroneous logic or inaccurate information. SophiaVL-R1 \cite{fan2025sophiavlr1reinforcingmllmsreasoning} was among the first to apply a think reward in MLLMs reasoning, achieving promising results. This leads to an intuitive hypothesis: \textit{Can a \textbf{think reward} that emphasizes the \textbf{thinking process} guide LALMs to improve their reasoning?}

To investigate this concept, we propose a model-generated think reward. This approach enables us to evaluate the nuanced reasoning quality of LALMs and examine their effects on final inference outcomes. We incorporate the Qwen3-8B-Base model as the think reward model, which assigns a score ranging from 0 to 1 in increments of 0.1 based solely on the quality of intermediate reasoning, independent of the correctness of the final answer. In instances where responses stem from the no-think mode, the think reward is calculated as the average of the think rewards within the batch.

\subsubsection{Overall Reward}
While integrating the consistency reward with other rewards can yield a high overall score even for incorrect answers, applying it exclusively when the final answer is correct mitigates undue emphasis on consistency. The think reward, in contrast, targets improvements in reasoning quality by evaluating intermediate steps, irrespective of the final answer’s correctness. The final reward structure is defined as follows.
\begin{equation}
R = R_{a} \times \left(1 + 0.5 \times R_{c}\right) +  0.5 \times R_{f}+R_{t}.
\end{equation}

\subsection{Reinforcement Learning }
Following DeepSeek-R1 \cite{shao2024deepseekmath}, given an input question $q$, GRPO samples a group of responses $\{o_1, o_2, \cdots, o_G\}$  , and their corresponding rewards corresponding rewards $\{R_1, R_2, \cdots, R_G\}$ are computed using the reward model. The advantage is subsequently computed as:
\begin{equation}
    \hat{A}_{i, t} = \widetilde{R}_i = \frac{R_i- {\rm mean}(\mathbf{R})}{{\rm std}(\mathbf{R})}
\end{equation}
The policy model is subsequently optimized by maximizing the Kullback-Leibler objective:
\begin{equation} \label{eq:GRPO_objective_revised}
\begin{split}
    &\mathcal{J}_{GRPO}(\theta) = \mathbb{E}_{\mathcal{D}} \Bigg[\frac{1}{G}\sum_{i=1}^G\frac{1}{|o_i|} \sum_{t=1}^{|o_i|} \Bigg\{ \min \Bigg[ \rho_{i,t} \hat{A}_{i,t}, \\
    &\quad \left. \operatorname{clip} \left( \rho_{i,t}, 1 - \epsilon, 1 + \epsilon \right)  \hat{A}_{i,t} \right] - \beta \mathbb{D}_{KL}\left[\pi_{\theta} || \pi_{ref}\right]\Bigg\}\Bigg]
\end{split}
\end{equation}
where $\rho_{i,t} = \frac{\pi_\theta(o_{i,t} | q, o_{i,<t})}{\pi_{\theta_{old}}(o_{i,t} | q, o_{i,<t})}$ is the probability ratio between the current policy $\pi_\theta$ and the policy $\pi_{\theta_{old}}$ , and $\epsilon$ and $\beta$ are hyper-parameters introduced in Proximal Policy Optimization (PPO) \cite{Schulman_Wolski_Dhariwal_Radford_Klimov_2017}.
\vspace{-2mm}
\section{Experiment}
\subsection{Experiment Setup}
\subsubsection{Dataset}
The training data is drawn from the AVQA dataset \cite{10.1145/3503161.3548291}, designed for audio-visual question answering and widely used in multimodal understanding research. Follow R1-AQA, we extract audio from videos and construct audio-text pairs by replacing ``video'' with ``audio'' in the questions, resulting in 40,176 training samples. For SFT with CoT, we first generate audio captions using Qwen2-Audio-7B-Instruct on AVQA. We then employ Qwen2.5-72B-Instruct\footnote{\url{https://huggingface.co/Qwen/Qwen2.5-72B-Instruct }} \cite{qwen2} to generate CoT rationales from the caption, question, and answer. The prompt used for CoT generation is provided in the Appendix \ref{CoT}.
\subsubsection{Implementation Details}
We use Qwen2-Audio-7B-Instruct and Qwen2.5-Omni as the basic models for experiments. The training is conducted on the SWIFT \cite{zhao2025swiftascalablelightweightinfrastructure} framework. To train our models, we use a node with 8 H20 GPUs (96GB). The batch size per GPU is 1 with gradient accumulation steps of 2 for a total effective batch size of 16. We train for 1000 steps on AVQA. We use a learning rate of 1e-6, a temperature of 1.0, 8 responses per GRPO step, and a KL coefficient $\beta$ of 0.04.
\subsection{Evaluation Metrics}
We evaluate model performance primarily by accuracy on multi-choice questions. Three main evaluation sets are used:

\begin{itemize}
    
    \item \textbf{MMAU Benchmark}: We evaluate the model using the test-mini set of the MMAU benchmark, which presents complex audio question-answer pairs that demand expert-level reasoning. Accuracy is determined by the percentage of correctly answered multiple-choice questions. The results of the officially updated MMAU benchmark, version v05.15.25\footnote{\url{https://sakshi113.github.io/mmau_homepage/}}, are provided in Appendix \ref{appendix:mmau}.
    
    \item \textbf{MMAR Benchmark}: This benchmark assesses deep reasoning across a range of real-world audio scenarios, incorporating mixed sounds, music, and speech, with questions specifically designed to challenge reasoning abilities. 
    
    \item \textbf{AIR Benchmark}: We analyze the model’s audio comprehension using the foundational sections of AIR-Bench, which encompasses a variety of audio modalities, including sound, speech, and music.
\end{itemize}

\begin{table*}[t!]
\centering
\setlength{\tabcolsep}{4pt} 

\scalebox{0.60}{
\begin{tabular}{cccccccccccc}
\toprule

\multirow{3}{*}{\textbf{Model}} & \multirow{3}{*}{\textbf{Method}} & & \multicolumn{4}{c}{\textbf{MMAU Test-mini}} & \multicolumn{4}{c}{\textbf{MMAR}} &\textbf{AIR} \\
\cmidrule(lr){4-7} \cmidrule(lr){8-11} \cmidrule(lr){12-12}
& && Sound$\uparrow$ & Music$\uparrow$ & Speech$\uparrow$ & Average$\uparrow$  & Sound$\uparrow$ & Music$\uparrow$ & Speech$\uparrow$ & Average$\uparrow$ & Average$\uparrow$ \\
\midrule

 Qwen2-Audio-7B-Instruct (reproduce) & - & &62.16&53.59&48.59&54.90 & 33.33 & 24.27 & 32.31 & 30.00&61.3  \\
\midrule

sft-a& SFT  &&63.66&56.59&54.35&58.20& 52.73 & 37.86 & 49.32 & 48.90 &63.8 \\
sft-b& SFT+CoT& & 63.36 & 56.29 & 54.41 & 57.80 & 56.36 &41.75 &48.30 &49.80&62.6\\
 \midrule
 
 grpo-a & GRPO  && 68.47 & 62.87 & 60.06 & 63.80 &56.36 &39.81 &48.98&50.20 &64.5\\
  grpo-b& GRPO+CoT&  & 70.27 & 63.17 & 61.56 & 65.00 &\textbf{58.18}&35.44&52.04&50.00 &64.1\\
 \midrule
 
model-a &  GRPO+ATAR &  & 74.47 & 63.47 & \underline{62.76} & 66.90 &\underline{57.58}&\textbf{54.55}&54.17&50.70&66.4 \\

 model-b  & GRPO +ATAR+ CR &  &\underline{74.77} & \underline{66.17} & 62.16 & \underline{67.70} &\textbf{58.18}&\underline{45.45}&\textbf{62.50}&\underline{50.90} &\underline{66.5} \\

model-c  &  GRPO +ATAR+ CR + TR & & \textbf{76.88} & \textbf{62.87} & \textbf{64.26} & \textbf{68.00} &56.97&\underline{45.45}&\underline{57.50}&\textbf{52.00}&\textbf{66.8} \\

\specialrule{0.15em}{0.3em}{0.3em}

 Qwen2.5-Omni (reproduce) & -  && 69.67 & 67.37 & 61.86 & 66.30 &61.21&49.51&57.14&58.20&64.9 \\
\midrule

sft-c& SFT & & \underline{77.18} & 62.57 & 63.96 & 67.90 & 63.03&50.00&57.82&60.90&65.8  \\
  sft-d& SFT+CoT &   & 75.98 & 63.47 & 63.06 & 67.50  &61.21&48.06&54.08&59.80&65.2 \\
 \midrule
 grpo-c & GRPO  && 75.38 & \underline{70.06} & 66.67 & 69.70  &66.06&51.94&62.24&62.50 &66.2\\
  grpo-d& GRPO+CoT & & 76.28 & 69.76 & 66.37 & 69.80&64.24&53.40&59.52&61.80 &65.9\\
 \midrule

model-d & GRPO+ATAR &  & 75.08 & 67.66 & 71.77 & 71.50& 63.64&\underline{54.85}&\underline{62.93}&64.20 &66.8\\

 model-e  & GRPO +ATAR+ CR & & 76.58 & 68.87 & \underline{72.07} & \underline{72.50}&\underline{66.67}&\textbf{55.83}&61.22&\underline{64.40} &\underline{67.0} \\

 model-f  &   GRPO+ATAR + CR + TR & & \textbf{77.48} & \textbf{70.36} & \textbf{73.37}& \textbf{73.70}&\textbf{67.27}&53.88&\textbf{64.29}&\textbf{65.30} &\textbf{67.1}\\

\bottomrule

\end{tabular}}
\caption{\textbf{Ablation Study Employing Qwen2-Audio-7B-Instruct and Qwen2.5-Omini as the Base Model}. The best-performing models in each category are highlighted in \textbf{bold}, and the second-best scores are \underline{underlined}. ATAR stands for Adaptive Think Accuracy Reward, CR stands for Consistency Reward, and TR stands for Think Reward.}
\label{Table1}
\end{table*}

\section{Results}
\subsection{Ablation Study }  
To systematically analyze the impact of different reasoning strategies and training methodologies, we conduct ablation studies using Qwen2-Audio-7B-Instruct and Qwen2.5-Omni as the baseline. Detailed experimental results are tabulated in Table \ref{Table1}.

\subsubsection{GRPO}  
We apply SFT and GRPO to Qwen2-Audio-7B-Instruct and Qwen2.5-Omni to develop several models: SFT (sft-a, sft-b, sft-c, sft-d) and GRPO (grpo-a, grpo-b, grpo-c, grpo-d).  GRPO models achieve significant improvements on the MMAU-test-mini, AIR Foundation, and MMAR benchmarks. However, explicit reasoning variants (grpo-b, grpo-d) do not outperform their implicit counterparts (grpo-a, grpo-c), suggesting that explicit reasoning alone provides insufficient guidance without effective supervision.

\subsubsection{Effectiveness of Adaptive Think Accuracy Reward}  
The comparison between model-a/d, which incorporate the adaptive thinking accuracy reward, and grpo-a/c and grpo-b/d, which are trained using the standard GRPO algorithm, highlights the effectiveness of the adaptive reward mechanism. Compared to grpo-a, the Qwen2-Audio-based model-a achieves improvements of 3.10, 0.50 and 1.9 in the MMAU-test-mini Avg, AIR Foundation Avg, and MMAR Avg, respectively. Compared to grpo-b, it shows gains of 1.90, 0.70, and 2.3 on the same metrics. Similarly, the Qwen2.5-Omni-based model-d outperforms grpo-c by 1.80, 1.70, and 0.6 on the three evaluation metrics, and shows improvements of 1.70, 2.40, and 0.9 over grpo-d. Collectively, these results indicate that the adaptive thinking accuracy reward enhances the model's reasoning performance.

\subsubsection{Necessity of Consistency Reward}  
The introduction of a consistency reward improves the performance of the model. Models incorporating the consistency reward (model-b/e) outperform those without it (model-a/d). Specifically, model-b achieves gains of 0.80, 0.20 and 0.10 over model-a on MMAU-test-mini Avg, AIR Foundation Avg, and MMAR Avg, respectively. Model-e shows improvements of 1.00, 0.20, and 0.20 across MMAU-test-mini Avg, AIR Foundation Avg, and MMAR Avg compared to model-d. This early reward stabilization mechanism effectively mitigates inconsistencies in the reasoning process.

\subsubsection{Impact of Think Reward}  
The integration of thinking rewards during reinforcement learning improves model performance. Models incorporating thinking rewards (model-c/f) consistently outperform those without the expert-LLM judging mechanism (model-b/e). Specifically, model-c achieves improvements of 0.30, 1.10, and 0.3 over model-b on MMAU-test-mini Avg, MMAR Avg, and AIR Foundation Avg, respectively. Similarly, model-f surpasses model-e by 1.20, 0.90, and 0.1 across the corresponding metrics. These results demonstrate the effectiveness of incorporating thinking rewards in guiding model learning.

\begin{table*}[htbp]
    \centering
    \scalebox{0.70}{
    \begin{tabular}{l rr rr rr rr rrrr }
        \toprule
         
        & \multicolumn{2}{c}{Sound} & \multicolumn{2}{c}{Music} & \multicolumn{2}{c}{Speech} & \multicolumn{2}{c}{Avg} \\
        \cmidrule(lr){2-3} \cmidrule(lr){4-5} \cmidrule(lr){6-7} \cmidrule(lr){8-9}
        Name & Test-mini & Test & Test-mini & Test & Test-mini & Test & Test-mini & Test \\
        \midrule
         
        Random Guess & 26.72 & 25.73 & 24.55 & 26.53 & 26.72 & 25.50 & 26.00 & 25.92 \\
        Most Frequent Choice & 27.02 & 25.73 & 20.35 & 23.73 & 29.12 & 30.33 & 25.50 & 26.50 \\
        Human (Test-Mini) & 86.31 & - & 78.22 & - & 82.17 & - & 82.23 & - \\
        \midrule
        GPT-4o Audio  \cite{jaech2024openai}& 61.56 & 56.27 & 56.29 & 55.27 & 66.37 & 67.20 & 61.40 & 59.58 \\
       Gemini  2.5 Flash \cite{comanici2025gemini25pushingfrontier} & 67.96 & 65.43 &62.28  & 65.30 & 62.76 &  63.30& 64.30 &64.68  \\
       \midrule
         \multicolumn{9}{c}{Pretrained + Supervised Finetuned Models} \\
        GAMA 7B \cite{ghosh2024gamalargeaudiolanguagemodel}& 41.44 & 45.40 & 32.33 & 30.83 & 18.91 & 19.21 & 30.90 & 31.81 \\
        Qwen Audio \cite{chu2023qwenaudioadvancinguniversalaudio} & 55.25 & 56.73 & 44.00 & 40.90 & 30.03 & 27.95 & 43.10 & 41.86 \\
        Qwen2 Audio \cite{qwen2} & 62.16 & 45.90 & 53.59 & 53.26 & 48.59 & 45.90 & 54.90 & 52.50 \\
        Mellow \cite{mellow} & 61.26 & 64.90 & 54.19 & 52.67 & 29.73 & 38.77 & 48.40 & 52.11 \\

        Audio Flamingo 2 \cite{audioflamingo2audiolanguage} & 61.56 & 65.10 & \underline{73.95} & \textbf{72.90} & 30.93 & 40.26 & 55.48 & 59.42 \\
        Kimi-Audio \cite{team2025kimi} & 61.68 & - & 73.27 & - & 60.66 & - & 65.00 & - \\
        
        \midrule
         \multicolumn{9}{c}{Finetuned with Reinforcement Learning} \\
          
         SARI (Qwen2-Audio) \cite{wen2025saristructuredaudioreasoning} & 68.55 & - & 69.01 & -& 59.09 & - & 65.55 & - \\
         SARI (Qwen2.5-Omni) \cite{wen2025saristructuredaudioreasoning} & 72.75 & - & 67.22 & -& 61.26 & - & 67.08 & - \\
        Audio-Reasoner \cite{xie2025audio} & 60.06 & - & 64.30 & - & 60.70 & - & 61.71 & - \\
        Audio-CoT \cite{ma2025audio} & 61.86 & - & 56.29 & - & 55.26 & - & 57.80 & - \\
        R1-AQA \cite{li2025reinforcement} & 68.77 & 69.76 & 64.37 & 61.40 & 63.66 & 62.70 & 65.60 & 64.36 \\
        Qwen2.5-Omni-7B \cite{xu2025qwen25omnitechnicalreport} & 69.67 & 70.63 & 67.37 & 66.93 & 61.86 & 66.57 & 66.30 & 68.03 \\
       
        AUDSEMTHINKER-QA GRPO \cite{wijngaard2025audsemthinkerenhancingaudiolanguagemodels} & 69.67 &69.20 & 69.16 & 63.13 & 61.26  & 65.77 & 66.70  & 66.03 \\
       
        Omni-R1 (VGGS-GPT) \cite{rouditchenko2025omnir1reallyneedaudio} & 73.6 &74.1 & \textbf{74.3} & \underline{70.8} & \underline{66.1}  & \underline{68.7}& \underline{71.3}  & \underline{71.2} \\
        \midrule
        \textsc{Audio-Thinker Qwen2-Audio} \textit{(ours)} & \underline{76.88} &\underline{75.13}  & 62.87 & 61.83 & 64.26  &67.03& 68.00 &67.90 \\
        \textsc{Audio-Thinker Qwen2.5-Omni} \textit{(ours)} & \textbf{77.48} & \textbf{76.30} & 70.36 & 66.63 & \textbf{73.37}& \textbf{73.27} & \textbf{73.70} & \textbf{72.83} \\
        \bottomrule
    \end{tabular}
    }
    \caption{Accuracy (\%) comparison on MMAU. For baselines, we evaluate GPT-4o Audio, Gemini 2.0 Flash, and Gemini 2.5 Flash. The results of other previous work are sourced from the original papers or the MMAU Leaderboard (old version).}
    \label{Table2}
\end{table*}

\begin{table*}[t!]
    \centering
    \scalebox{0.9}{
    \begin{tabular}{l r r rrr r}
        \toprule
        & \multicolumn{1}{c}{AIR-Sound} & \multicolumn{1}{c}{AIR-Music} & \multicolumn{3}{c}{AIR-Speech} & AIR-Avg \\
       \cmidrule(lr){2-2} \cmidrule(lr){3-3} \cmidrule(lr){4-6} \cmidrule(lr){7-7}
        Model & SoundAQA & MusicAQA & SER & VSC & SNV & Avg \\
        \midrule
        Gemini 2.0 Flash \cite{narzary2025comparativestudyzeroshotcrosslingual} 
            & 69.9 & 68.2 & 56.2 & 93.5 & 64.8 & 66.1 \\
        Gemini 2.5 Flash \cite{comanici2025gemini25pushingfrontier} 
            & 74.8 & \textbf{73.7} & 56.4 & 94.1 & \textbf{68.5} & \textbf{67.4} \\
        GPT-4o Audio \cite{jaech2024openai} 
            & 68.3 & 67.7 & 51.2 & 90.0 & 61.6 & 62.3 \\
        \midrule
        SALMONN \cite{qwen2} 
            & 28.4 & 54.6 & 29.9 & 45.3 & 34.3 & 36.8 \\
        Minmo \cite{minmo} 
            & 50.3 & - & \textbf{64.5} & 93.0 & - & - \\
        Qwen2-Audio-Instruct \cite{qwen2} 
            & 67.2 & 64.6 & 50.5 & 87.9 & 60.5 & 61.3 \\
        Qwen2.5-Omni-7B \cite{xu2025qwen25omnitechnicalreport} 
            & 75.3 & \underline{70.6} & 56.4 & 92.9 & 63.9 & 64.9 \\
        Audio-Reasoner \cite{xie2025audio} 
            & 65.7 & 55.2 & \underline{60.5} & - & 56.3 & 65.2 \\
        \midrule
        \textsc{Audio-Thinker Qwen2-Audio} \textit{(ours)}   
            & \underline{75.5} & 68.7 & 55.7 & \underline{94.4} & 64.5 & 66.8 \\
        \textsc{Audio-Thinker Qwen2.5-Omni} \textit{(ours)} 
            & \textbf{75.8} & 69.5 & 56.2 & \textbf{94.5} & \underline{67.5} & \underline{67.1} \\
        \bottomrule
    \end{tabular}}
    \caption{Accuracy (\%) comparison on AIR foundation and MMAR. For baselines, we evaluate GPT-4o Audio, Gemini 2.0 Flash, and Gemini 2.5 Flash on the AIR-Bench foundation. We obtain the reported results for other previous work from their original papers and the AIR paper.}
    \label{Table3}
\end{table*}

\begin{table*}[t!]
    \centering
    \scalebox{0.9}{
    \begin{tabular}{l r rrr r}
        \toprule
        & \multicolumn{4}{c}{MMAR}  \\
       \cmidrule(lr){2-5} \cmidrule(lr){6-6}
        Model & Sound & Music & Speech & Avg  \\
        \midrule
        Gemini 2.0 Flash \cite{narzary2025comparativestudyzeroshotcrosslingual} 
            & 61.21 & 50.97 & 72.11 & 65.20 \\
        Gemini 2.5 Flash \cite{comanici2025gemini25pushingfrontier} 
            & 55.28 & \underline{53.40} & \textbf{77.21} & \textbf{66.80}  \\
        GPT-4o Audio \cite{jaech2024openai} 
            & 53.94 & 50.97 & 70.41 & 63.50  \\
        \midrule
        SALMONN \cite{qwen2} 
            & 30.91 & 29.61 & 24.35 & 32.80 \\
        Qwen2-Audio-Instruct \cite{qwen2} 
            & 33.33 & 24.27 & 32.31 & 30.00  \\
        Qwen2.5-Omni-7B \cite{xu2025qwen25omnitechnicalreport} 
            & 58.79 & 40.78 & 59.86 & 56.70  \\
        Audio-Reasoner \cite{xie2025audio} 
            & 43.64 & 33.50 & 32.99 & 36.80  \\
        Omni-R1 (VGGS-GPT) \cite{rouditchenko2025omnir1reallyneedaudio} 
            & \underline{67.3} & 51.5 & 64.3 & 63.4  \\
        \midrule
        \textsc{Audio-Thinker Qwen2-Audio} \textit{(ours)}   
            & 56.97 & 45.45 & 57.50 & 52.00  \\
        \textsc{Audio-Thinker Qwen2.5-Omni} \textit{(ours)} 
            & \textbf{68.32} & \textbf{53.88} & 64.29 & \underline{65.30}  \\
        \bottomrule
    \end{tabular}}
    \caption{Accuracy (\%) comparison on  MMAR. For baselines, we evaluate Gemini 2.5 Flash on MMAR. We obtain the reported results for other previous work from their original papers and the MMAR paper. Detailed results are presented in the Appendix \ref{sec:mmar}.}
    \label{Table4}
\end{table*}
\subsection{Compare with SOTA }
\subsubsection{MMAU}
Table \ref{Table2} summarizes the key results from the MMAU benchmark. For baseline models, we highlight recently proposed methods that have achieved SOTA performance. Notably, compared to the Qwen2.5-Omni baseline, Audio-Thinker (Qwen2.5-Omni) improves the average performance on Test-mini from 66.30 to 73.70, and on Test-full from 68.03 to 72.83. Compared to the Qwen2-Audio baseline, Audio-Thinker (Qwen2-Audio) also shows substantial improvements, with Test-mini performance increasing from 54.90 to 68.00, and Test-full performance rising from 52.50 to 67.90. Among all previously reported models, Audio-Thinker (Qwen2.5-Omni) achieves the highest scores in both the sound and speech categories, as well as in the overall average, performing exceptionally well on both the Test-mini and Test-full datasets. Notably, compared to the previous SOTA Omni-R1 model, which is also based on Qwen2.5-Omni, our model achieves absolute improvements of 2.40 and 1.63 in Test-mini Avg and Test-full Avg, respectively.

\subsubsection{AIR}
Table \ref{Table3} presents results from the AIR-Bench foundation benchmark, which evaluates audio understanding across three primary categories: sound, music, and speech. The speech category is further divided into three subdomains: Speech Emotion Recognition (SER), Vocal Sound Classification (VSC), and Speech Number Variation (SNV). In terms of the overall AIR-Bench foundation average, Audio-Thinker (Qwen2.5-Omni) achieves 67.1, outperforming all existing open-source models and even surpassing several closed-source systems including GPT-4o Audio \cite{jaech2024openai}, though it remains behind the most powerful Gemini 2.5 Flash \cite{comanici2025gemini25pushingfrontier} model.

In the sound category, Audio-Thinker (Qwen2.5-Omni) scores 75.8 and Audio-Thinker (Qwen2-Audio) scores 75.5, outperforming Audio-Reasoner (65.7) and Qwen2.5-Omni (75.3), setting a new benchmark. In music reasoning, Audio-Thinker (Qwen2.5-Omni) scores 69.5, slightly below Qwen2.5-Omni (70.6). In speech reasoning, Audio-Thinker (Qwen2.5-Omni) scores 56.2 in SER, 94.5 in VSC (highest overall), and 67.5 in SNV (second-best score). Its exceptional performance in speaker recognition reinforces its strengths in speech tasks.
\subsubsection{MMAR}
 Table \ref{Table4} summarizes the results from the MMAR evaluation. We focus on Qwen2-Audio and Qwen2.5-Omni as baseline models, with additional comparative results available in the respective original studies. Notably, Audio-Thinker (Qwen2.5-Omni) outperforms all existing open-source models, including Omni-R1, which is based on the same Qwen2.5-Omni architecture but trained on a larger dataset. This demonstrates the effectiveness of the Audio-Thinker framework in enhancing deep audio reasoning. Furthermore, our models achieve performance levels comparable to, and in some cases surpassing, those of current SOTA closed-source systems such as Gemini 2.5 Flash and GPT-4o Audio, as illustrated at the top of the Table \ref{Table3}. These results provide strong evidence that Audio-Thinker effectively improves the deep reasoning capabilities of LALMs.

\section{Conclusion}
\label{sec:conclusion}

In this work, we present Audio-Thinker, an audio-language reinforcement learning framework that integrates model-generated think-based rewards with adaptive outcome rewards. This approach guides the model towards difficulty-aware, consistent, and effective reasoning. To enhance adaptive reasoning, we introduce an adaptive thinking accuracy reward, allowing the model to modify its reasoning strategy according to the task's complexity. Additionally, we tackle the issue of reward hacking by incorporating think-based rewards that assess the quality of the reasoning process. Experimental results across various benchmarks reveal that Audio-Thinker consistently outperforms existing LALMs. Our findings underscore the significance of adaptive reasoning and the importance of supervising the thinking process beyond mere final correctness, providing valuable insights for the future development of audio-language reasoning models.

\clearpage
\bibliography{main}

\appendix
\newpage

\appendix

\section*{\hspace{-4mm} \centering Appendix}
\vspace{3mm}
\section{Prompt Details}
\subsection{Prompt of Adaptive Think\label{appendix:adaptive}}
\begin{tcolorbox}[width=1.0\textwidth]
   First, identify whether this problem requires
thinking. If the problem requires thinking,
output thinking process in <think> </think> and
final answer inside <answer> </answer>. If no
thinking is required, and the final output
answer in <answer> </answer>
The Assistant is encouraged to use the <answer></answer> tag whenever possible, while ensuring accuracy.
\end{tcolorbox}
\subsection{Prompt of Generating CoT Dataset\label{CoT}}
\begin{tcolorbox}[width=1.0\textwidth]
   We are developing a system to generate structured audio-based chain-of-thought reasoning data. Given an audio clip, its description, a question, and an answer, your task is to reconstruct the reasoning process in two parts: the internal <think> section and the user-facing <response> section. The <think> section must follow four steps: planning, captioning, reasoning, and summarizing. Based on this, the <response> should provide a final answer. Your output must strictly follow this format: <think><planning>. Analyze the user’s intent and break down complex tasks into steps if needed. </planning><caption> Examine the audio input, identify relevant segments, and describe them accurately. </caption><reasoning> Use the identified information to reason toward an answer. </reasoning><summary> Conclude based on the reasoning above. </summary></think><answer> Give the final answer here referring to the <think> part </answer> \\
Please strictly follow the format of the sample.\\

Note that you have both the question and the answer because it is necessary to ensure the correctness of the chain of thought. However, in your response, you can only refer to the content of the question and the audio, which leads to the answer. You must not assume that you already know the answer.\\

Here is the original description: \textbf{*** caption here ***}. \\

The question is: \textbf{*** question here ***}. \\

The answer you can refer to: \textbf{*** answer here ***}. \\
\end{tcolorbox}
\subsection{Think Prompt}
\begin{tcolorbox}[width=1.0\textwidth]
   Please first think about the reasoning process in the mind, and then provide the user with the action. The reasoning process and answer are enclosed within <think> </think> and <answer> </answer> tags, respectively, i.e., <think> reasoning process here </think><answer> answer here </answer> 
\end{tcolorbox}

\newpage

\section{More Experiments}

\subsection{MMAU (v05.15.25)}\label{appendix:mmau}

The official updated release of the MMAU benchmark, MMAU-v05.15.25, incorporates valuable feedback from the research community, with approximately 25\% of the questions and answers revised to improve clarity, precision, and overall quality, and around 5\% of the audio files refined to enhance acoustic consistency and signal fidelity, establishing this version as a stable and improved reference for future evaluation. The official MMAU leaderboard has been updated to reflect the performance of state-of-the-art models on MMAU-v05.15.25. 

Due to the official MMAU benchmark not yet being updated during our preliminary experiments—particularly in the iterative process of developing the Audio-Thinker framework by incorporating multiple reward components—we adopted the previous version of MMAU (old version). To ensure experimental consistency, all results presented in the main text are based on this prior version. Following the release of the updated MMAU-v05.15.25 benchmark, we further evaluated our trained Audio-Thinker models on this new version to assess their performance on the latest benchmark and to enable direct comparison with current state-of-the-art models.

As shown in Table~\ref{tab:mmau_table}, Audio-Thinker trained models on MMAU-v05.15.25 achieve significant improvements over the baseline, consistent with the performance trend observed on the original MMAU benchmark. Moreover, compared to previous state-of-the-art models, Audio-Thinker (Qwen2.5-Omni) achieves the best performance to date, establishing a new SOTA on the updated benchmark.

\begin{table*}[htbp]
    \centering
    \resizebox{\linewidth}{!}{
    \begin{tabular}{l rr rr rr rr}
        \toprule
        & \multicolumn{2}{c}{Sound} & \multicolumn{2}{c}{Music} & \multicolumn{2}{c}{Speech} & \multicolumn{2}{c}{Avg} \\
        \cmidrule(lr){2-3} \cmidrule(lr){4-5} \cmidrule(lr){6-7} \cmidrule(lr){8-9}
        Name & Test-mini & Test & Test-mini & Test & Test-mini & Test & Test-mini & Test \\
        \midrule
        \multicolumn{9}{c}{\textbf{closed-models}} \\
        GPT-4o Audio  \cite{jaech2024openai} & 64.56 & 63.20 & 56.29 & 49.93 & 66.67 & 69.33 & 62.50 & 60.82 \\
        GPT-4o mini Audio  \cite{jaech2024openai} & 50.75 & 49.67 & 39.22 & 35.97 & 69.07 & 67.47 & 53.00 & 51.03 \\
        Gemini 2.0 Flash \cite{comanici2025gemini25pushingfrontier} & 71.17 & 68.93 & 65.27 & 59.30 & 75.08 & 72.87 & 70.50 & 67.03 \\
        Gemini 2.5 Flash \cite{comanici2025gemini25pushingfrontier} & 73.27 & 69.50 & 65.57 & 69.40 & 76.58 & 68.27 & 71.80 & 67.39 \\
        Gemini 2.5 Pro \cite{comanici2025gemini25pushingfrontier} & 75.08 & 70.63 & 68.26 & 64.77 & 71.47 & 72.67 & 71.60 & 69.36 \\
        Gemini 2.5 Flash Lite \cite{comanici2025gemini25pushingfrontier} & 63.06 & 62.50 & 63.47 & 54.87 & 72.07 & 67.47 & 66.20 & 61.61 \\
        \midrule
        \multicolumn{9}{c}{\textbf{Pretrained + Supervised Finetuned Models}} \\
        GAMA 7B \cite{ghosh2024gamalargeaudiolanguagemodel} & 31.83 & 30.73 & 17.71 & 17.33 & 12.91 & 16.97 & 20.82 & 21.68 \\
        Qwen2-Audio-Instruct \cite{qwen2} & 67.27 & 61.17 & 56.29 & 55.67 & 55.26 & 55.37 & 59.60 & 57.40 \\
        Audio Flamingo 2 \cite{audioflamingo2audiolanguage} & 61.56 & 65.10 & 73.95 & \textbf{72.90} & 30.93 & 40.26 & 55.48 & 59.42 \\
        Audio Flamingo 3 \cite{goel2025audioflamingo3advancing} & - & 76.67 & - & 73.30 & - & 64.87 & - & 72.28 \\
        Kimi-Audio \cite{team2025kimi} & 75.68 & 70.70 & 66.77 & 65.93 & 62.16 & 56.57 & 68.20 & 64.40 \\
        Step-Audio 2 \cite{stepaudio2} & 77.4 & - & \textbf{82.0} & - & 75.7 & - & 74.6 & - \\
        \midrule
        \multicolumn{9}{c}{\textbf{Finetuned with Reinforcement Learning}} \\
        Audio-Reasoner \cite{xie2025audio} & 67.87 & 67.27 & 69.16 & 61.53 & 66.07 & 62.53 & 67.70 & 63.78 \\
        Qwen2.5-Omni-7B \cite{xu2025qwen25omnitechnicalreport} & 78.10 & 76.77 & 65.90 & 67.33 & 70.60 & 68.90 & 71.50 & 71.00 \\
        Omni-R1 (FT on Qwen2.5-Omni) \cite{rouditchenko2025omnir1reallyneedaudio} & \underline{81.7} & \underline{78.3} & 73.4 & \underline{70.8} & \underline{76.0} & \underline{75.8} & \underline{77.0} & \underline{75.0} \\
        \midrule
        \textsc{Audio-Thinker Qwen2-Audio} \textit{(ours)} & 77.48 & 74.67 & 65.57 & 63.83 & 67.57 & 67.06 & 70.20 & 68.52 \\
        \textsc{Audio-Thinker Qwen2.5-Omni} \textit{(ours)} & \textbf{82.58} & \textbf{79.03} & \underline{74.55} & 70.53 & \textbf{76.88} & \textbf{76.60} & \textbf{78.00} & \textbf{75.39} \\
        \bottomrule
    \end{tabular}
    }
    \caption{\textbf{Performance Comparison on MMAU-v05.15.25.} Results for other methods are sourced from the MMAU Leaderboard: MMAU-v05.15.25. The best-performing models in each category are highlighted in \textbf{bold}, and the second-best scores are \underline{underlined}.}
    \label{tab:mmau_table}
\end{table*}

\newpage

\subsection{MMAR}\label{sec:mmar}

MMAR \cite{ma2025mmarchallengingbenchmarkdeep} is a rigorously designed evaluation benchmark aimed at probing the deep reasoning capabilities of contemporary Audio-Language Models (ALMs) across a broad spectrum of real-world acoustic scenarios. Specifically, MMAR comprises 1,000 audio–question–answer triplets collected from open-domain internet videos, which are then refined through a multi-stage process involving expert annotation, iterative error correction, and statistical quality control to ensure high data fidelity and ecological validity.

In contrast to extant benchmarks that confine evaluation to narrow taxonomic silos—namely isolated speech, music, or generic sound events—MMAR adopts a pan-spectrum design philosophy, explicitly incorporating mixed-modality compositions (e.g., sound–music, sound–speech, music–speech, and fully overlapped sound–music–speech) to reflect the polyphonic complexity of everyday auditory scenes. Each query is hierarchically stratified across four ascending reasoning layers: Signal Layer: low-level acoustic attribute extraction (e.g., bandwidth, SNR, spectral centroid); Perception Layer: mid-level perceptual grouping and event boundary detection; Semantic Layer: high-level conceptual mapping and cross-modal entailment; Cultural Layer: sociocultural or pragmatic inference grounded in auditory context.

Table \ref{table6} empirically situates our proposed Audio-Thinker family—anchored on Qwen2-Audio and Qwen2.5-Omni backbones—within the broader landscape of previous work. Quantitative results demonstrate consistent superiority across all four reasoning layers, corroborating the efficacy of our paradigm.

\begin{table}[htbp]
\centering
\resizebox{\linewidth}{!}{
\begin{tabular}{l c c c c c c c c c}
\toprule
\multirow{2}{*}{\textbf{Models}} & \multirow{2}{*}{\textbf{Size}} & \multicolumn{3}{c}{\textbf{Single Modality (\%)}} & \multicolumn{4}{c}{\textbf{Mixed Modalities (\%)}} & \multirow{2}{*}{\textbf{Avg (\%)}} \\
\cmidrule(lr){3-5} \cmidrule(lr){6-9}
 & & \textbf{Sound} & \textbf{Music} & \textbf{Speech} & \textbf{Sound-Music} & \textbf{Sound-Speech} & \textbf{Music-Speech} & \textbf{Sound-Music-Speech} & \\
\midrule
Random Guess & - & 29.39 & 25.88 & 31.48 & 25.00 & 29.30 & 31.10 & 28.13 & 29.32 \\
\midrule
\multicolumn{10}{c}{\textbf{Large Audio Language Models (LALMs)}} \\
Audio Flamingo~\cite{audio-flamingo} & 2.2B & 32.73 & 21.84 & 24.83 & 18.18 & 30.28 & 24.39 & 25.00 & 26.60 \\
Audio Flamingo 2~\cite{audioflamingo2audiolanguage} & 0.5B & 20.61 & 20.39 & 24.15 & 27.27 & 23.85 & 26.83 & 25.00 & 23.00 \\
\midrule
\multicolumn{10}{c}{\textbf{Audio-Thinker}} \\
\textsc{Audio-Thinker Qwen2-Audio} & 7B & 56.97 & 45.63 & 57.50 & 36.36 & 47.71 & 48.78 & 62.50 & 52.00 \\
\textsc{Audio-Thinker Qwen2.5-Omni} & 7B & \textbf{68.48} & \textbf{53.88} & 64.29 & 72.73 & 71.56 & 73.17 & 66.67 & 65.30 \\
\bottomrule
\end{tabular}}
\caption{
\textbf{MMAR results across six model categories: LALMs, LARMs, OLMs, LLMs, LRMs with audio captions as input, and our Audio-Thinker models}. The results for prior models are sourced from the original MMAU\cite{ma2025mmarchallengingbenchmarkdeep} paper and their respective original publications. The best-performing models in each category are highlighted in \textbf{bold}, and the second-best ones are \underline{underlined}.
}
\label{table6}
\end{table}

\newpage

\section{More information about the dataset}\label{sec:dataset_more}
The training data is derived from the AVQA dataset \cite{yang2022avqa}, a rigorously curated resource originally conceived for joint audio-visual comprehension in unconstrained, real-world video environments. The AVQA dataset comprises 57,015 high-resolution video recordings depicting quotidian human activities—ranging from domestic chores to urban traffic scenes—accompanied by 57,335 human-annotated question–answer pairs that systematically probe object–action relations, spatio-temporal causality, and cross-modal semantic alignment.
To adapt this corpus for audio-only reasoning, we conduct a modality-specific re-formulation pipeline:
\textbf{Modal Isolation}. We extract the monophonic audio streams (16 kHz, 16-bit PCM) from each video while discarding visual frames.
\textbf{Lexical Re-contextualization}.  Each original question is automatically parsed and surface-normalized so that deictic references to “video” or “frame” are replaced with “audio” or “segment”, ensuring linguistic coherence within an audio-centric schema.
\textbf{Quality Filtering and Validation}.  We apply heuristics (SNR > 10 dB, min. duration 1.5 s) and human spot-checks to retain only \textbf{40,176} high-fidelity audio–question pairs—approximately 70\% of the original training split—thereby yielding a balanced, audio-grounded reasoning corpus that preserves the original relational diversity while eliminating visual redundancy.
This re-purposed dataset serves as the primary pre-training substrate for our Large Audio Reasoning Models, enabling them to learn fine-grained acoustic discrimination, temporal event parsing, and causal inference in the absence of visual cues.

\section{Hyperparameter Configuration}\label{sec:hyperparameter}
\begin{tcolorbox}[width=1.0\textwidth]
Rlhf\_type: GRPO \\
Train\_type: lora \\
Lora\_rank: 8 \\
Lora\_alpha: 32 \\
Target\_modules: all-linear \\
Torch\_dtype: bfloat16 \\
Max\_completion\_length: 2048 \\
Max\_steps: 1000 \\
Per\_device\_train\_batch\_size: 1 \\
Per\_device\_eval\_batch\_size: 1 \\
Learning\_rate: 1e-6 \\
Gradient\_accumulation\_steps: 2 \\
Warmup\_ratio: 0.05 \\
Num\_generations: 8 \\
Temperature: 1.0 \\
Top\_p: 0.99 \\
Top\_k: 50 \\
Deepspeed: zero3 \\
\end{tcolorbox}

\appendix
\end{document}